\documentclass[sigconf,screen]{acmart}

\acmConference[ICSE 2022]{The 44th International Conference on Software Engineering}{May 21–29, 2022}{Pittsburgh, PA, USA}

\usepackage{graphicx}
\usepackage[many]{tcolorbox}

\usepackage{mathtools,amssymb,latexsym,amsfonts,stmaryrd}
\usepackage{pbox}
\usepackage{xfrac}
\usepackage{booktabs}
\usepackage{xcolor}
\usepackage{threeparttable}
\usepackage{amsthm}
\usepackage{hyperref}
\usepackage{multirow}
\usepackage{tikz}
\usepackage{mathrsfs}
\usepackage{mathpartir}
\usepackage{algorithm}
\usepackage{url}
\usepackage{syntax}
\usepackage{framed}
\usepackage[noend]{algpseudocode}
\usepackage{flushend}
\usepackage{listings}
\usepackage{moresize}
\usepackage{wrapfig}
 
\usepackage{seqsplit}
\usepackage{alltt}
\usepackage{xspace}
\usepackage{enumitem}
\usepackage{pifont}

\DeclareMathAlphabet{\mathcal}{OMS}{cmsy}{m}{n}

\usepackage{amsmath, listings, amsthm, amssymb, proof, xspace}

\usepackage{thmtools}

\declaretheoremstyle[spaceabove=\topsep,notefont=\normalfont\itshape]{mystyle}

\newcommand{\revise}[2]{{\color{red}{\ifx&#1&\else- #1\fi}} {\color{ForestGreen}{\ifx&#2&\else+ #2\fi}}}%
\renewcommand{\revise}[2]{#2}%

\usetikzlibrary{arrows,patterns, decorations.pathreplacing}

\newcommand{\F}{Fig.}

\newcommand{\T}{Table}
\renewcommand{\S}{Sec.}

\newcommand{\ignore}[1]{}

\lstdefinestyle{base}{
  moredelim=**[is][\color{red}]{@}{@},
  escapeinside={<@}{@>}
}

\newcommand{\tool}{\textsc{IRGen}\xspace}
\newcommand{\clang}{\texttt{clang}\xspace}
\newcommand{\gcc}{\texttt{gcc}\xspace}

\newcommand{\ncc}{\texttt{ncc}\xspace}
\newcommand{\ncci}{\texttt{ncc-w/o-inst2vec}\xspace}
\newcommand{\codebert}{\texttt{CodeBERT}\xspace}
\newcommand{\codecmr}{\texttt{CodeCMR}\xspace}
\newcommand{\ctv}{\texttt{code2vec}\xspace}
\newcommand{\cts}{\texttt{code2seq}\xspace}
\newcommand{\aroma}{\texttt{Aroma}\xspace}
\newcommand{\aromad}{\texttt{Aroma-Dot}\xspace}
\newcommand{\aromac}{\texttt{Aroma-Cos}\xspace}
\newcommand{\misim}{\texttt{MISIM}\xspace}
\newcommand{\misimg}{\texttt{MISIM-GNN}\xspace}

\usepackage{amssymb}
\usepackage{pifont}

\newcommand\DejaVuttfamily{%
  \fontfamily{DejaVuSansMono-TLF}\selectfont }

\lstdefinestyle{base}{
  moredelim=**[is][\color{red}]{@}{@},
  escapeinside={<@}{@>}
}

\lstdefinelanguage
   [x64]{Assembler}     
   [x86masm]{Assembler} 
   {morekeywords={CDQE,CQO,CMPSQ,CMPXCHG16B,JRCXZ,LODSQ,MOVSXD, %
                  POPFQ,PUSHFQ,SCASQ,STOSQ,IRETQ,RDTSCP,SWAPGS, %
                  rax,rdx,rcx,rbx,rsi,rdi,rsp,rbp, %
                  r8,r8d,r8w,r8b,r9,r9d,r9w,r9b}} 

\usepackage{listings}
\usepackage{fancyvrb}
\usepackage{color}
\definecolor{lightgray}{rgb}{.9,.9,.9}
\definecolor{darkgray}{rgb}{.4,.4,.4}
\definecolor{purple}{rgb}{0.65, 0.12, 0.82}
\definecolor{commentgreen}{RGB}{63,127,95}

\colorlet{myPurple}{blue!40!red}
\definecolor{myOrange}{RGB}{255,192,0}

\newcommand{\enc}[1]{$\phi^{*}_{\theta}$}
\newcommand{\dec}[1]{$\psi^{*}_{\theta}$}

\lstdefinelanguage{Solidity}{
  keywords={len,delete,int,void,payable, public, event, contract, typeof, new, true, false, catch, function, return, null, catch, switch, var, if, in, while, do, else, case, break,struct,const,socklen_t,sa_familty_t,char,sockaddr},
  keywordstyle=\color{violet}\bfseries,
  ndkeywords={class, export, boolean, throw, implements, import, this},
  ndkeywordstyle=\color{darkgray}\bfseries,
  identifierstyle=\color{black},
  sensitive=false,
  comment=[l]{//},
  escapeinside={(*@}{@*)},          
  morecomment=[s]{/*}{*/},
  commentstyle=\color{commentgreen}\ttfamily,
  stringstyle=\color{red}\ttfamily,
  morestring=[b]',
  morestring=[b]"
}

\newcommand{\rnum}[1]{\uppercase\expandafter{\romannumeral #1\relax}}

\usepackage{xcolor}

\algnewcommand{\LeftComment}[1]{\Statex \(\triangleright\) #1}

\definecolor{pptbrown}{RGB}{132,60,12}
\definecolor{pptgreen}{RGB}{56,87,35}

\let\OLDthebibliography\thebibliography
\renewcommand\thebibliography[1]{
  \OLDthebibliography{#1}
  \setlength{\parskip}{0pt}
  \setlength{\itemsep}{0pt plus 0.1ex}
}

\definecolor{pptgreen}{RGB}{84,130,53}
\definecolor{pptred}{RGB}{176,35,24}

\definecolor{pptgreen1}{RGB}{78,173,91}
\definecolor{pptred1}{RGB}{192,0,0}

\definecolor{pptyellow1}{RGB}{203,195,167}
\definecolor{pptgreen2}{RGB}{184,192,176}

\definecolor{pptblue}{RGB}{194,214,236}

\definecolor{pptgreen3}{RGB}{146,208,80}

\acmSubmissionID{icse22-main-2066}

\copyrightyear{2022}
\acmYear{2022}
\setcopyright{acmcopyright}\acmConference[ICSE '22]{44th International Conference on Software Engineering}{May 21--29, 2022}{Pittsburgh, PA, USA}
\acmBooktitle{44th International Conference on Software Engineering (ICSE '22), May 21--29, 2022, Pittsburgh, PA, USA}
\acmPrice{15.00}
\acmDOI{10.1145/3510003.3510217}
\acmISBN{978-1-4503-9221-1/22/05}

\begin{document}

\title{Unleashing the Power of Compiler Intermediate Representation to Enhance
  Neural Program Embeddings}

\author{Zongjie Li, Pingchuan Ma, Huaijin Wang, Shuai~Wang}
\affiliation{%
  \institution{The Hong Kong University of Science and Technology}
  \country{Hong Kong SAR}
}
\authornote{Corresponding author}
\email{{zligo,pmaab, hwangdz, shuaiw}@cse.ust.hk}

\author{Qiyi Tang, Sen Nie, Shi Wu}
\affiliation{%
  \institution{Tencent Security Keen Lab}
  \country{China}
}
\email{{dodgetang, snie, shiwu}@tencent.com}

\begin{abstract}

  Neural program embeddings have demonstrated considerable promise in a range of
  program analysis tasks, including clone identification, program repair, code
  completion, and program synthesis. However, most existing methods generate
  neural program embeddings directly from the program source codes, by learning
  from features such as tokens, abstract syntax trees, and control flow graphs.

  This paper takes a fresh look at how to improve program embeddings by
  leveraging compiler intermediate representation (IR). We first demonstrate
  simple yet highly effective methods for enhancing embedding quality by
  training embedding models alongside source code and LLVM IR generated by
  \textit{default} optimization levels (e.g., \texttt{-O2}). We then introduce
  \tool, a framework based on genetic algorithms (GA), to identify
  (near-)optimal sequences of optimization flags that can significantly improve
  embedding quality.
   
  We use \tool\ to find optimal sequences of LLVM optimization flags by
  performing GA on source code datasets. We then extend a popular code embedding
  model, \codecmr, by adding a new objective based on triplet loss to enable a
  joint learning over source code and LLVM IR. We benchmark the quality of
  embedding using a representative downstream application, code clone detection.
  When \codecmr\ was trained with source code and LLVM IRs optimized by findings
  of \tool, the embedding quality was significantly improved, outperforming the
  state-of-the-art model, \codebert, which was trained only with source code.
  Our augmented \codecmr\ also outperformed \codecmr\ trained over source code
  and IR optimized with default optimization levels.
  We investigate the properties of optimization flags that increase embedding
  quality, demonstrate \tool's generalization in boosting other embedding
  models, and establish \tool's use in settings with extremely limited training
  data. Our research and findings demonstrate that a straightforward addition to
  modern neural code embedding models can provide a highly effective
  enhancement.

\end{abstract}

\maketitle

\section{Introduction}
\label{sec:introduciton}

Recent developments in deep neural networks (DNNs) have delivered advancements
in computer vision (CV) and natural language processing (NLP) applications. We
have noticed lately an increase in interest in using DNNs to solve a variety of
software engineering (SE) problems, including software
repair~\cite{wang2018dynamic,gupta2017deepfix}, program
synthesis~\cite{matej2016deepcoder,liang2017neural,parisotto2016neurosymbolic},
reverse engineering~\cite{shin2015recognizing}, malware
analysis~\cite{ben2018neural}, and program
analysis~\cite{zhang2018neural,si2018nips}. Similar to how DNNs understand
discrete natural language text, nearly all neural SE applications require 
computing numeric and continuous representations over software, which are referred
to as program embeddings or embedding vectors.

The common procedure for generating code embeddings is to process a program's
source code directly, extracting token sequences, statements, or abstract syntax
trees (ASTs) to learn program representations~\cite{poj, gupta2017deepfix,
  bhatia2016automated,alon2019code2vec, alon2018code2seq}. Although some
preliminary approaches have attempted to extract semantics-level code
signatures, such approaches are limited by use of semantic features that are too
coarse-grained~\cite{piech2015learning}, low code coverage (due to dynamic
analysis)~\cite{wang2018dynamic}, or limited
scalability~\cite{wang2019learning}.
To date, learning from code syntactic and structural information has remained
the dominant approach in this field, and as previous work has
argued~\cite{wang2017dynamic,wang2019learning,sui2020flow2vec,ding2019asm2vec},
the use of features at this relatively ``shallow'' level is likely to degrade
learning quality and produce embeddings with low robustness.

For some CV and NLP tasks, \textit{data augmentation} has been proposed as a
tool to improve the quality of learned embedding
representations~\cite{shorten2019survey,feng2021survey}. These approaches
typically increase the amount of training data by adding slightly modified
copies of already existing data or by creating new pieces of synthetic data from
existing data. Thus, embedding models can be trained on larger numbers of data
samples, resulting in higher-quality embedding representations. Previous
research has shown the value of data augmentation approaches in increasing
embedding
quality~\cite{min2020syntactic,fawzi2016adaptive,li2019improving,lopes2019improving}.

This work investigates using compiler intermediate representations (IR) to
augment code embedding. Modern compilers include numerous optimization flags
that can seamlessly convert a piece of source code into a range of semantically
identical but syntactically distinct IR codes. From a comprehensive standpoint,
we argue that our technique can boost program embedding on two fundamental
levels. \textbf{First}, the translation of a single piece of source code into
several variants of IR code with the same functionality significantly increases
the diversity of available training data. As previously noted, such augmented
data can commonly improve the quality of learned embeddings. \textbf{Second},
although programs with the same functionality may appear syntactically distinct
as source code, they are likely to become more similar after pruning and
rearrangement by optimizations. This alleviates the difficulties imposed by
syntax changes, as the optimizations regulate syntactic characteristics.

We begin by illustrating that using default compiler optimization levels, such
as \texttt{-O2} of LLVM, can produce IR code that significantly improves the
embedding quality of a popular embedding model, \codecmr~\cite{yu2020codecmr},
and outperforms the state-of-the-art (SOTA) model,
\codebert~\cite{feng2020codebert}, trained on source code alone.
However, despite the promising potential in this ``misuse'' of compiler
optimizations, the high number of available optimization flags and the
consequently large search space present a challenge for identifying
well-performing optimization sequences to augment embedding models.

We propose \tool, a framework that uses genetic algorithms (GA) to search for
(near-)optimal optimization sequences for generation of IR code to augment
program embedding models. Compiler optimization flags are typically combined to
generate machine instructions with high speed or small size. In contrast, \tool\
targets optimization sequences, generating IR code that is \textit{structurally
similar} to the input source code. This prevents over-simplification of the IR
code, which is undesirable in our task since overly-simplified IR often becomes
less ``expressive.'' Additionally, to maximize learning efficiency, we limit
overuse of out-of-vocabulary (OOV) terms (our definition of OOV follows
\ncc~\cite{ncc}; see \S~\ref{subsec:fitness}).

We present a simple yet unified extension, through triplet
loss~\cite{weinberger2009distance}, to enable embedding models to learn from
source code and LLVM IR. For evaluation, we used \tool\ to analyze IRs generated
from the POJ-104~\cite{poj} and GCJ~\cite{gcj} datasets, which include a total
of 299,880 C/C++ programs. After 143 to 963 CPU hours of search (we use a
desktop computer to run \tool), \tool\ was able to form optimization sequences
with high fitness scores from the 196 optimization flags available in the x86
LLVM framework (ver.~11.1.0). 
To evaluate the quality of embedding, we setup a representative downstream task,
code clone detection. When \codecmr\ was trained with IR code generated by the
identified optimization sequences, embedding quality (in terms of code clone
detection accuracy) significantly improved by an average of 11.66\% (peaking at
15.46\%), outperforming the SOTA model \codebert\ trained with only source code
(for 12.02\%) or \codecmr\ jointly trained with source code and IR emitted by
default optimizations (for 5.94\%).
%
We also demonstrate that \tool\ is general to augment other neural embedding
models and show that \tool\ can almost \textit{double} the quality of learned
embeddings in situations with limited data (e.g., 1\% of training data
available). We characterize optimization flags selected by \tool\ and summarize
our findings. This work can help users take use of compiler optimization, an
\textit{out-of-the-box} amplifier, to improve embedding quality. In summary, our
contributions are as follows:

\begin{itemize}
\item We advocate the use of compiler optimizations for software embedding
  augmentation. Deliberately-optimized IR code can principally improve the
  quality of learned program embeddings by extending model training datasets and
  normalizing syntactic features with modest cost.

\item We build \tool, a practical tool that uses GA algorithms to iteratively
  form (near-)optimal optimization sequences. Additionally, we present a simple
  yet general extension over modern code embedding models to enable joint
  learning over source code and IR.

\item Our evaluation demonstrates highly promising results, with our augmented
  model significantly outperforming SOTA models. We further demonstrate the
  generalization of \tool\ and its merit in augmenting very limited training
  data. \tool\ is released at~\cite{website}.
\end{itemize}

\section{Preliminary}
\label{sec:preliminary}

\begin{figure}[!ht]
  \centering
  \includegraphics[width=1.00\linewidth]{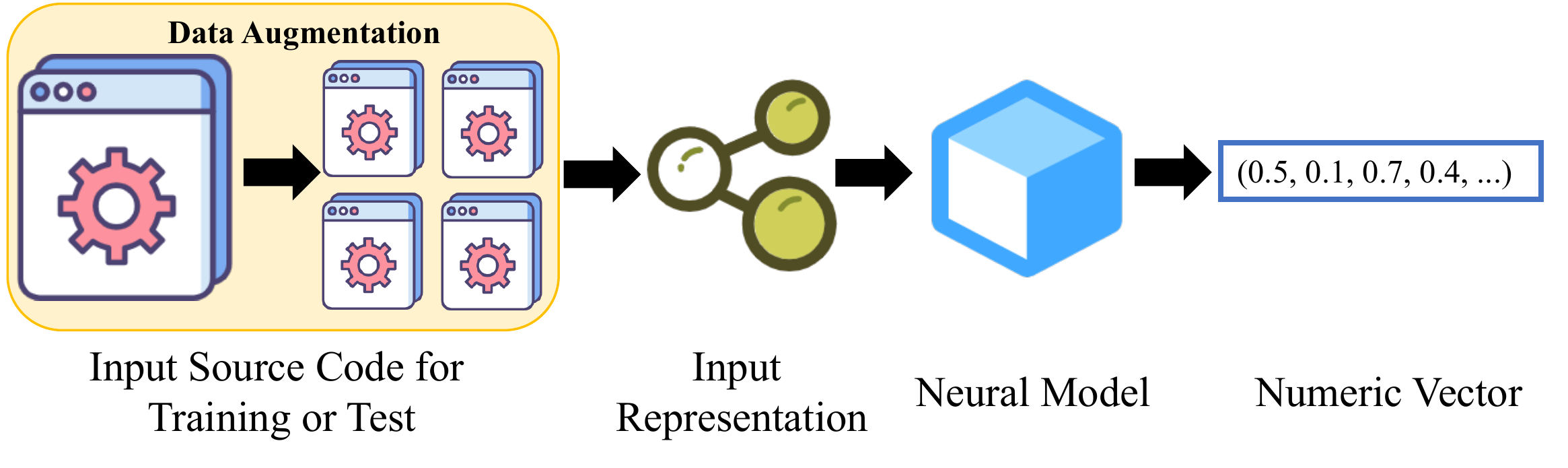}
  \caption{Common neural program embedding pipeline.}
  \label{fig:pipeline}
\end{figure}

Neural code embedding, as in \F~\ref{fig:pipeline}, converts discrete source
code to numerical and continuous embedding vectors, with the end goal of
facilitating a variety of learning-based program analysis. We
introduce program representations in \S~\ref{subsec:background-input}. We
examine alternative model designs in \S~\ref{subsec:background-model} and the
concept of data augmentation for neural (software) embedding in
\S~\ref{subsec:background-augmentation}.

\subsection{Input Representation}
\label{subsec:background-input}

Code can be expressed as text and processed using existing NLP models. However,
it would be costly and likely ineffective because programming languages usually
contain a wealth of explicit and sophisticated structural information that is
difficult for NLP models to comprehend~\cite{poj}. Therefore, modern code
embedding models often learn program embeddings using code structural
representations which are informative. For instance, the abstract syntax tree
(AST) is used to represent code fragments for program embeddings. Once a code
snippet's AST has been generated, there are several methods for extracting
discrete symbols (e.g., AST nodes) for use in the subsequent learning process.
For example, \ctv~\cite{alon2019code2vec} and \cts~\cite{alon2018general}
extract a collection of paths from AST to form embeddings, as discussed below.

Control flow graphs (CFG) are also used to form input representation, especially
when analyzing assembly code. Two representative tools,
asm2vec~\cite{ding2019asm2vec} and BinaryAI~\cite{yu2020order}, construct CFGs
over assembly code and combine basic block-level embeddings into the program's
overall embedding. Recent
research~\cite{allamanis2018learning,guo2020graphcodebert,ncc} has explored the
use of hybrid representations that incorporate data from different layers. For
instance, \ncc~\cite{ncc} extracts a so-called contextual flow graph first,
which subsumes information from both control flow graph and data flow graph.

\subsection{Neural Model Learning Procedure}
\label{subsec:background-model}

NLP models are generally designed to process infinite sequences of tokens,
whereas software is structured. Hence, the neural code embedding learning
process can be divided into two broad categories: 1) decomposing program
(structural) representations (e.g., AST or CFG) into one or multiple token
sequences that are then processed by NLP models; and 2) attempting to initiate
an ``end-to-end'' procedure for directly learning structural representations
using advanced neural models like graph neural networks (GNN).

\codebert\ is a large-scale SOTA code embedding model that primarily learns from
token-level software representations. It is inspired by
BERT~\cite{devlin2018bert}, a famous bidirectional natural language embedding
model. \codebert\ constructs learning objectives using both masked language
modeling (MLM) and replacement token detection. Using these objectives, it is
trained to predict tokens that have been randomly masked out of the inputs until
saturation accuracy is reached. Another model,
\texttt{asm2vec}~\cite{ding2019asm2vec}, uses MLMs, particularly an extended
PV-DM model~\cite{le2014distributed}, to embed x86 instructions at the token
level. Token sequences can be extracted from tree or graph representations. For
example, \ctv~\cite{alon2019code2vec} and \cts~\cite{alon2018code2seq} break AST
into paths, transform paths to embeddings using LSTM~\cite{hochreiter1997long},
and finally aggregate path-level embeddings to produce the AST's embedding. The
structure-based traversal method~\cite{hu2018deep} converts ASTs into structured
sequences.

TBCNN~\cite{mou2014tbcnn}, Great~\cite{hellendoorn2019global} and
BinaryAI~\cite{yu2020order} leverage advanced models, such as GNNs, to directly
process program structural representations. BinaryAI~\cite{yu2020order}, for
example, uses standard GNNs to propagate and aggregate basic block embeddings
into CFG embeddings. Besides CFGs, neural models can create structures with
richer information. \ncc~\cite{ncc} forms a contextual flow graph with control-
and data-flow information. Each node in the contextual flow graph contains a
list of LLVM IR statements, which \ncc\ then transforms into vectors. It further
uses a GNN to aggregate the node embeddings into an embedding of the entire
program.
As with \ncc, \misim\ begins by constructing a novel context-aware semantic
structure (CASS) from collections of program syntax- and structure-level
properties. It then converts CASS into embedding vectors using GNNs. It
outperforms prior AST-based embedding tools, including \ctv\ and
\texttt{code2seq}~\cite{alon2018code2seq}.

\subsection{Data Augmentation}
\label{subsec:background-augmentation}

Images can be rotated while retaining their ``meaning'' (e.g., via affine
transformations~\cite{zhang2015character}). Similarly, we can replace words in
natural language sentences with their synonyms, which should not impair
linguistic semantics. Data augmentation leverages these observations to create
transformation rules that can enlarge model training data.

It is worth noting a conventional technique, namely feature
engineering~\cite{zheng2018feature}, can generally help data science and machine
learning tasks. Feature engineering facilitates to eliminate redundant data that
can reduce overfitting and increase accuracy. Nevertheless, in the era of deep
learning, it gradually becomes less desirable to manually ``pick useful
features,'' given that we need to frequently deal with high-dimensional data
like image, text, video, and software. How to pick useful features is often
obscure when learning from those complex high-dimensional data. In fact, it has
been demonstrated that data augmentation generally and notably improves deep
learning model performance and robustness, and it has been frequently employed
as a routine technique to enhance modern deep learning models in a variety of
domains~\cite{zhang2015character, wei2019eda, ma2020metamorphic, ribeiro2019red,
selvaraju2020squinting, wang2020metamorphic, pang2021mdpfuzzer,
yuan2021enhancing, ma2022mtteql, yuan2021perception}.

Standard data augmentation approaches, however, are \textit{not} directly
applicable to enhance program embeddings. Augmenting neural program embeddings
is challenging and under-explored. Due to the synthetic and semantic constraints
of programming languages, arbitrary augmentation can easily break a well-formed
program.
This paper explores bringing data augmentation to source code. In particular, we
advocate employing compiler optimizations to turn a same piece of source code
into semantically identical but syntactically diverse IR code. Note that we do
not need to ``reinventing the wheel'' to develop extra semantics-preserving
source code transformations~\cite{jain2021contrastive}. Instead, we demonstrate
how a mature compiler can facilitate effective data augmentation simply by
exploiting optimizations developed over decades by compiler engineers. 


\section{Motivation}
\label{sec:motivation}

The LLVM compiler architecture supports hundreds of optimization passes, each of
which mutates the compiler IR in a unique way. To make compiler optimization
more accessible to users, the LLVM framework offers several optimization bundles
that a user can specify for compilation, for example, \texttt{-O2},
\texttt{-O3}, and \texttt{-Os}. The first two bundles combine optimization
passes for fast code execution, whereas \texttt{-Os} aims to generate the
smallest executable possible.
%
Our preliminary study shows that by incorporating optimized IR code into
embedding learning, the embedding quality can be substantially enhanced. This
section describes our preliminary finding, which serves as an impetus for the
subsequently explored research.

\begin{table}[t]
  \centering
  \scriptsize
  \caption{MAP scores of \codecmr\ on POJ-104~\cite{poj} for different input setup.}
  \label{tab:opt-comparation}
  \resizebox{1.00\linewidth}{!}{
    \begin{tabular}{cccc}
      \hline
      \textbf{Setup}  & MAP(\%)   & \textbf{Setup}  & MAP(\%)  \\
      \hline
      \texttt{Source}                  &76.39 & \texttt{Source + LLVM IR -O2} &84.29 \\
      \texttt{Source + LLVM IR -O0}    &82.90 & \texttt{Source + LLVM IR -O3} &84.21 \\
      \texttt{Source + LLVM IR -O1}    &83.37 & \texttt{Source + LLVM IR -Os} &83.81 \\
      \hline
      \multicolumn{3}{l}\texttt{\hspace{-5pt}Source + LLVM IR Optimized by Optimization Sequences Found by \tool}  &  89.18  \\
      \hline
    \end{tabular}
  }
\end{table}

\noindent \textbf{Learning Over Source Code.}~We use POJ-104~\cite{poj}, a
commonly used dataset containing 44,912 C programs written for 104 tasks. This
dataset is split into three program sets: one for training, one for validation,
and one for testing (see \S~\ref{sec:evaluation}). We trained
\codecmr~\cite{yu2020codecmr}, one popular code embedding tool, on the training
split, and then perform multi-label classification on the testing split.
\codecmr\ generates code embeddings by first converting source code to a
character sequence and then computing character-level embeddings. The embeddings
are fed to a stack of Pyramid Convolutional Neural Network
(DPCNN)~\cite{johnson2017deep}, in which an average pooling layer constructs the
program's embedding. DPCNN has been shown as powerful at embedding programs. In
our evaluation on POJ-104, we observe promising accuracy in terms of
MAP~\cite{MusgraveBL20} score, as shown in \texttt{Source} of
\T~\ref{tab:opt-comparation}. MAP is a commonly used metrics in this field, and
a higher MAP score indicates a greater quality of code embeddings. As will be
shown in evaluation (\T~\ref{tab:map}), this result is comparable to those
obtained using the SOTA model, \codebert.

\noindent \textbf{Findings.}~Despite the decent results, we find that
\codecmr\ fails to group quite a number of POJ-104 programs who belong to the
same class. The POJ-104 C code and corresponding LLVM IR are too lengthy to fit
in the paper; we present some very readable cases at~\cite{llvmir} and summarize
our key observations below.

C/C++ programs implementing the same functionality can exhibit 
distinct syntactic appearance. For instance, at~\cite{llvmir}, we present a case
where two programs, $p_1$ and $p_2$, are implementing the same ``days between
dates'' task. We find that $p_1$ uses one \texttt{switch} statement, whereas
$p_2$ uses a sequence of \texttt{if} statements. Further, $p_1$ uses many local
variables to encode \#days in each month, while those information in $p_2$ are
hardcoded in constants. This way, $p_1$ and $p_2$, differ
from both control- and data-flow perspectives.

Nevertheless, we find that the LLVM IR code compiled from these two programs are
much closer in both control- and data-flows. Let $l_1$ and $l_2$
(see~\cite{llvmir}) be two LLVM IR programs compiled from $p_1$ and $p_2$ and
optimized with optimization level -O3. We find that $l_1$ and $l_2$ preserves
most of the structure of the source code. More crucially, $l_1$ and $l_2$ both
use a LLVM IR \texttt{switch} statement to encode the control structures. Data
usage is also regulated, where both local variables in $p_1$ and the constants
in $p_2$ become integers hardcoded in IR statements. The induced IR programs
$l_1$ and $l_2$ are (visually) very similar, revealing the true semantics-level
equivalence of $p_1$ and $p_2$. We thus suspect that \codecmr\ is indeed
hampered by too flexible code representation in C programs. In other words, it
is shown as demanding to explore extracting more robust features from C/C++ code
to enhance the learning quality.

\noindent \textbf{Learning over Code Structure or Semantics.}~As previously stated,
\codecmr\ learns on the character (token) sequence. This indicates that
\codecmr\ is less resilient against changes at the syntactic level. Graph-level
embeddings might be more robust to token-level changes, given their reliance on
the rich structural information contained in the program. Nonetheless, in
real-life code samples, many changes can also occur at the graph level, and as
shown in \S~\ref{sec:evaluation}, representative graph-level embedding models
also perform poorly on diverse and large-scale datasets, such as POJ-104.

Some readers may wonder if learning directly from \textit{code semantics}, such
as input-output behaviors captured by dynamic
analysis~\cite{wang2017dynamic,wang2019learning}, is possible. While dynamic
analysis can precisely describe code behaviors (on the covered paths), it
suffers from low coverage. Symbolic execution (SE)~\cite{cadar2008klee} is used
to include a greater amount of program activity in applications such as code
similarity analysis~\cite{luo2014semantics}. Nonetheless, SE is inherently
inefficient, where trade-offs are made to launch SE over real-world
software~\cite{cadar2015targeted,trabish2018chopped}.

\noindent \textbf{Learning over IR Code.}~This paper advocates using compiler IR to
extend model train dataset and enhance code embedding models. However, we do
\textit{not} suggest learning solely from IR for two reasons. First, compiler
optimizations such as static single-assignment
(SSA)~\cite{cytron1991efficiently} result in LLVM IR codes that typically have
ten times as many lines of code (LOC) as the corresponding C source code. This
provides a significant impediment to training embedding models. In our
preliminary study, we find that training embedding models using LLVM IR code
alone resulted in significantly inferior performance across multiple tasks and
datasets.
Second, when outputting IR code, the LLVM compiler prunes or ``obfuscates''
certain source code features such as string and variable names. Note that
variable names and constants are generally crucial to improving embedding
quality. Similarly, in LLVM IR code, callsites, particularly to standard
libraries like \texttt{glibc}, are often modified. For example, the callsite
statement in \texttt{set<int> row; row.insert(x);} would be converted to a
complex function name with additional prefixes. Notably, we should avoid
tweaking with or disabling certain ``annoying'' optimization passes (for example,
the SSA pass), as many optimization flags assume the existence of other flags.

\noindent \textbf{Learning Over Source Code and IR Code.}~We extended
\codecmr\ to process IR code emitted by \clang. We augmented the frontend of
\codecmr\ to process LLVM IRs. We also extended the learning objectives by
requiring \codecmr\ to minimize the distance between the source code and 
corresponding IR using triplet loss (see \S~\ref{subsec:loss}). We compiled
each test program in the POJ-104 training set into LLVM IR to train the
\codecmr, and then benchmark the MAP score using the same setting.

As seen in \T~\ref{tab:opt-comparation}, using LLVM IR in the learning process
significantly improved embedding performance. For instance, when compiling the
source code into LLVM IR with negligible optimization (\texttt{-O0}), the joint
learning enhanced the MAP score by approximately 6\%. Note that jointly training
over source code and IR (\texttt{-O0}) has already outperformed the SOTA model,
\codebert\ (82.7\%). More importantly, it is seen that compiler optimizations
can notably improve \codecmr's performance. We observe that as compared with
\texttt{-O0}, using optimization levels \texttt{-O2}, \texttt{-O3}, and
\texttt{-Os} produces MAP scores greater than 84\%.
%

We regard the above findings as encouraging and intuitive: they demonstrate the
possibility and benefit of learning jointly from source code and IR code (which
are more regulated) rather than from source code alone. In evaluation
(\S~\ref{subsec:rq3}), we discuss optimization flags further with case studies
to reveal that they can effectively regulate code generation patterns, remove
superfluous code fragments, and generate more consistent IR code in the presence
of syntactic or structural changes in the source code. We therefore summarize
the key findings of this early investigation as follows:

\begin{tcolorbox}
  Launching a joint training using both program source code and 
  corresponding IR can notably improve the embedding quality.
\end{tcolorbox}
%

\noindent \textbf{Limitation of Standard Optimization Levels.}~Despite the
encouraging results, we note that these default optimization levels are selected
by compiler engineers with \textit{different} focus, e.g., producing smallest
possible executable or fast execution.
However, we explore a different angle, where optimizations are ``misused'' to
generate LLVM IR code to \textit{augment program neuron embedding}. In that
regard, it is possible to suspect the inadequacy of utilizing simply the default
optimization levels. For instance, certain CPU flags in \texttt{-Os} and
\texttt{-O3} are aggressive in shrinking IR code (e.g.,
\texttt{-aggressive-instcombine}), which, might not be proper since embedding
models generally prefer more ``expressive'' inputs. In evaluation, we find that
aggressive flags such as \texttt{-aggressive-instcombine} are not picked by
\tool. We also find that optimization flags should be adaptive to source code of
different complexity, whereas default optimization levels are fixed.
\S~\ref{subsec:rq3} compares optimization flags selected by \tool\ when
analyzing different datasets.

We introduce \tool, an automated framework to determine (near-)optimal sequences
of optimization flags for each particular dataset. To compare with standard
optimization levels, the final row of \T~\ref{tab:opt-comparation} presents the
improved performance of \codecmr\ when using \tool-selected optimization flags.
These results show that IR code optimized using \tool-formed optimization
sequence significantly improved the accuracy.



\begin{figure}[!ht]
  \centering
  \includegraphics[width=1.0\linewidth]{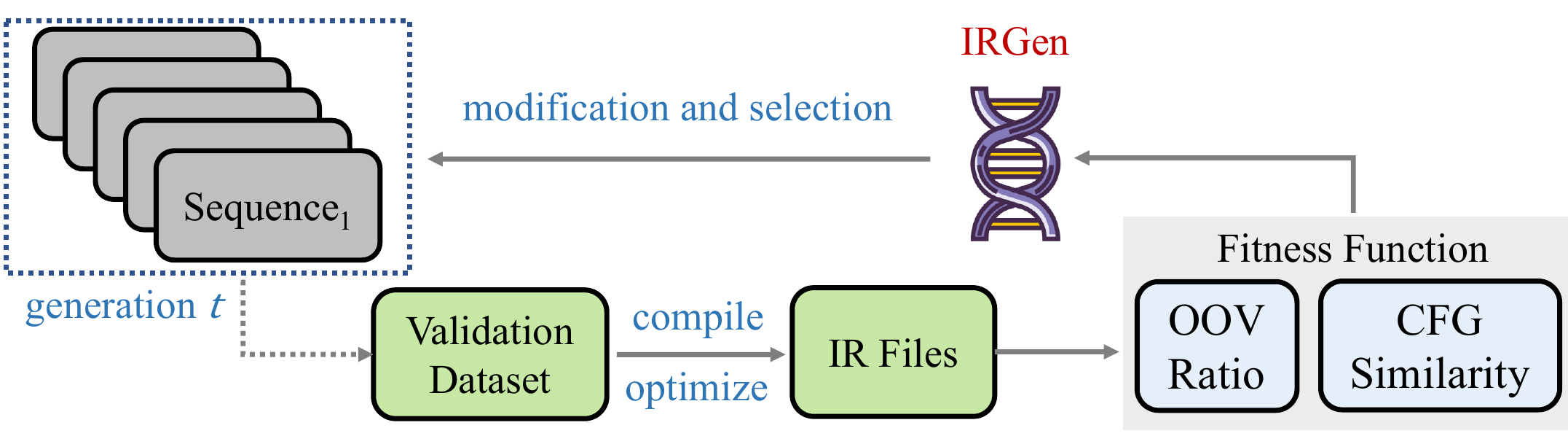}
  \caption{The workflow of \tool.}
  \label{fig:workflow}
\end{figure}


\section{Design of \tool}
\label{sec:design}


\F~\ref{fig:workflow} depicts \tool's workflow. Given a dataset of C source
code, we first form a validation dataset $P$ where fitness scores are computed
from (see \S~\ref{subsec:fitness}). Our observation shows that the size of $P$
does not need to be large (otherwise the GA procedure becomes slow). For the
current implementation, we randomly select 5\% programs from the training
dataset (POJ-104 or GCJ; see details in \S~\ref{sec:evaluation}) of \codecmr.

\tool\ initializes the first generation of optimization sequences
(\S~\ref{subsec:state-action}), and then launches the GA-based search to
iteratively modify and select sequences giving high fitness scores
(\S~\ref{subsec:action}). The GA process is repeated $N$ times until termination
($N$ is 800 currently). After termination, we select the $K$ sequences with the
top-$K$ highest fitness scores across the entire GA search. Using these
optimization sequences, each piece of C/C++ code in the training data can be
compiled into $K$ syntactically distinct pieces of LLVM IR code. The resulting
augmented code dataset can be used to empower neural embedding models, by
incorporating triplet loss as an extra learning objective
(\S~\ref{subsec:loss}).

\noindent \textbf{Application Scope.}~This research mainly focuses on the LLVM
compiler framework given its popularity. LLVM provides a total of 196
optimization flags that are applicable on x86 platforms. \tool\ traverses the
entire search space to identify sequences expected to improve embedding quality.
\tool's iterative method is \textit{orthogonal} to the LLVM framework and can
therefore be extended to support other compilers, such as \gcc\ (to manipulate
its GIMPLE IR), without incurring additional technical difficulties. 

The current implementation of \tool\ primarily enhances C/C++ code embedding.
C/C++ programs can be compiled into LLVM IR and optimized accordingly. Moreover,
most security related embedding downstream applications (e.g., CVE
search~\cite{ding2019asm2vec}) concern C/C++ programs. Nevertheless, we clarify
that code embedding has its wide application on other languages such as
Java/Python. It is worth noting that \tool\ relies on a rich set of compiler
optimizations to generate diverse IR code. Java/Python compilers/interpreters
provide fewer optimization passes, and they leave many optimizations at runtime.
As a result, the search space for \tool\ to explore would be much smaller. We
also have a concern on the expressiveness of Java/Python bytecode in comparison
with LLVM IR. Their bytecode seems very succinct, potentially undermining the
SOTA embedding models. Overall, we leave it as one future work to explore
extending \tool\ to enhance Java/Python embedding. 

\tool's present implementation does not consider the order of optimizations in a
sequence. We also assume each flag can only be used once. This enables a
realistic and efficient design when GA is used; similar design decisions are
also made in relevant
works~\cite{liu2017stochastic,ren2021unleashing}.
Taking orders or repeated flags into account would notably enlarge the search
space and enhance the complexity of \tool. We reserve the possibility of using
metaheuristic algorithms with potentially greater capacity, such as deep
reinforcement learning~\cite{mnih2013playing}, for future work. See
\S~\ref{sec:discussion} for further discussion.

\noindent \textbf{Design Focus.}~\tool's GA-based pipeline was inspired by
literatures in search-based software engineering, particularly using GA for code
testing, debugging, maintenance, and hardening~\cite{mcminn2011search,
  alba2007acohg, o2008search, ghaith2012improving, wang2017composite,
  ren2021unleashing}. \S~\ref{sec:related} further reviews existing studies. Our
evaluation will show that the GA method, when combined with our well-designed
fitness function, is sufficiently good at selecting well-performing optimization
sequences. Further enhancement may incorporate other learning-based techniques;
see \S~\ref{sec:discussion}.

\subsection{Genetic Representation}
\label{subsec:state-action}


Following common GA practice, we represent each optimization sequence as a
one-dimensional vector $v=(f_1, f_2, \ldots, f_L)$, where $L$ is the total
number of optimization flags offered by LLVM for x86 platforms. Each $f_i$ is a
binary number (0/1), denoting whether the corresponding flag, $c_i$, is enabled
or not on sequence $i$.
As standard setup, we initialize $M$ instances of vector $v$, by randomly
setting elements in a vector $v$ as 1. These randomly initialized sets, referred
to as a ``population'' in GA terminology, provide a starting point to launch
generations of evolution. Here, $M$ is set as 20.

\subsection{Modification and Selection}
\label{subsec:action}



At each generation $t$, we employ two standard genetic operators, Crossover and
Mutation, to manipulate all 20 vectors in the population. Given two ``parent''
vectors, $v_1$ and $v_2$, two offsprings are generated using $k$-point
crossover: $k$ cross-points are randomly selected on $v_1$ and $v_2$, and the
content marked by each pair of cross-points is swapped between them. Here, $k$
is set as 2, and the chance of each potential crossover is set as 0.4. We also
use flip bit mutation, another common method, to diversify vectors in the
population. We randomly mutate $1\%$ of bits in vector $v$. After these
mutations, the population size remains unchanged (20 vectors), but some vectors
are modified. Each vector is assessed using the fitness function defined in
\S~\ref{subsec:fitness}. All mutated and unmutated vectors are then passed into
a standard roulette wheel selection (RWS) module, where the chance for selecting
a vector is proportional to its fitness score. This way, a vector with a higher
fitness score is more likely to be selected into the next generation. The RWS
procedure is repeated 20 times to prepare 20 vectors for generation $t+1$.


\subsection{Fitness Function}
\label{subsec:fitness}

Given a vector, $v$, denoting a sequence of optimization flags, fitness function
$\mathcal{F}$ yields a fitness score as an estimation of $v$'s merit.
Specifically, for each $v$, we compile every program $p$ in the validation
dataset $P$ using optimizations specified in $v$ to produce IR programs $l \in
L$. For a C program $p$ and its compiled IR $l$, we first compute the following
fitness score:

\[
 Fitness\_Score_{p,l} = sim_{G} \times \frac{unk\_rate_{0}}{unk\_rate_{l}}
\]

\noindent where $sim_{G}$ denotes the graph-level similarity between the $l$ and
$p$. The value of $unk\_rate_{0}$ denotes the number of \#OOV cases found in IR
code $l_0$ when compiling $p$ with \texttt{-O0} (i.e., the baseline), and
$unk\_rate_{l}$ stands for \#OOV cases found in $l$. Then, $\mathcal{F}$ is
acquired by averaging the above fitness score for all programs $p \in P$.

\noindent \textbf{Graph Similarity.}~The graph similarity metric quantifies the
similarity between the original source code and the compiled IR code at the CFG
level. This provides a high-level assessment of the created IR code's quality.
More importantly, this condition prevents excessive reduction of the code by the
compiler optimizations, ensuring that the IR code reasonably preserves the
original source code’s structure-level features.

We tentatively assessed three graph similarity computation methods: 1) kernel
methods, 2) graph embedding~\cite{wu2020comprehensive,goyal2018graph}, and 3)
tree edit distance. Graph embedding methods often require to fine-tune a large
number of hyper-parameters which is generally challenging. 
We also find that tree edit distance algorithms had limited capacity to process
the very complicated CFGs created for our test cases. \tool's present
implementation therefore uses a classic and widely-used kernel method,
shortest-path kernels~\cite{borgwardt2005shortest}, to quantify the structural
distances between source code $p$ and its optimized IR code $l$. Overall, kernel
methods, including shortest-path kernels, denote a set of methods originated
from statistical learning theory to support pattern
analysis~\cite{campbell2002kernel}. Kernel methods are shown to be effective in
various tasks such as classification and regression. For our scenario, we feed
the employed kernel function with a pair of CFG derived from source code $p$ and
the corresponding IR $l$, and the kernel function returns a score $sim_{G}$.

\noindent \textbf{OOV Ratio.}~In embedding learning, OOVs represent tokens that
are rarely observed and are not part of the typical token vocabulary. We clarify  
that we follow \ncc~\cite{ncc} to define vocabulary. Particularly, our
vocabulary denotes a bag of IR statements, and therefore, IR code is represented
by a list of statement embeddings. Accordingly, ``OOV'' in our context denotes a
new statement never occurring in the baseline vocabulary. Such new statements
correspond to a special embedding noted as ``[unknown]'' in our implementation,
degrading the learning quality.

A high \#OOV is discouraged in the fitness function. That is, we leverage the
OOV ratio to punish an optimization sequence if it results in an IR code with
too many OOV cases. To this end, a ``baseline'' is first computed, recording the
\#OOV encountered in IR code generated by compiling $p$ with \texttt{-O0}. Then,
given optimization sequence $v$, we count the \#OOV cases identified in its
optimized IR code $l$, and compute the relative OOV ratio.

We clarify that it is possible to avoid token-level OOV issue by leveraging
sub-tokenization techniques like BPE~\cite{karampatsis2020big}. Given that said,
in the current setting, an IR statement is represented by a single embedding
vector, whereas BPE represents a statement by multiple vectors of sub-tokens.
The extra overhead due to multiple vectors is seen as acceptable for source code
but unaffordable for IR code, which is orders of magnitude longer. In fact, our
preliminary study explored using BPE: we report that BPE would result in 16$\times$ and
30$\times$ longer vectors on our test datasets, POJ-104~\cite{poj} and GCJ~\cite{gcj}.

\subsection{Learning Multiple IRs using Triplet Loss}
\label{subsec:loss}

Instead of keeping single sequence with the highest fitness score,
\tool\ retains the top-$K$ sequences from each generation, as ranked by their
fitness scores. We find that it is beneficial to perform augmentation with
multiple LLVM IR codes generated by the top-$K$ optimization sequences (see
results in \S~\ref{sec:implementation}). Given the GA procedure, these top-$K$
sequences will evidently share some overlapping optimization flags. However, we
find that when a source program is compiled into $K$ LLVM IR programs using
these top-$K$ sequences, these $K$ IR programs are still distinct (see cases
at~\cite{diverseir}), although they share regulated code structures that are
correlated with the reference source code. Hence, we anticipate that the
augmented dataset will be diverse, which has been generally shown to be useful
in enhancing embedding learning
quality~\cite{Cameron2019dataaug,Varun2020dataaugtr,LiuXJMWV20}. $K$ denotes a
hyper-parameter of \tool. We benchmark the accuracy changes in terms of
different $K$ in \S~\ref{sec:implementation}.

\begin{figure}[!ht]
  \centering
  \includegraphics[width=1.0\linewidth]{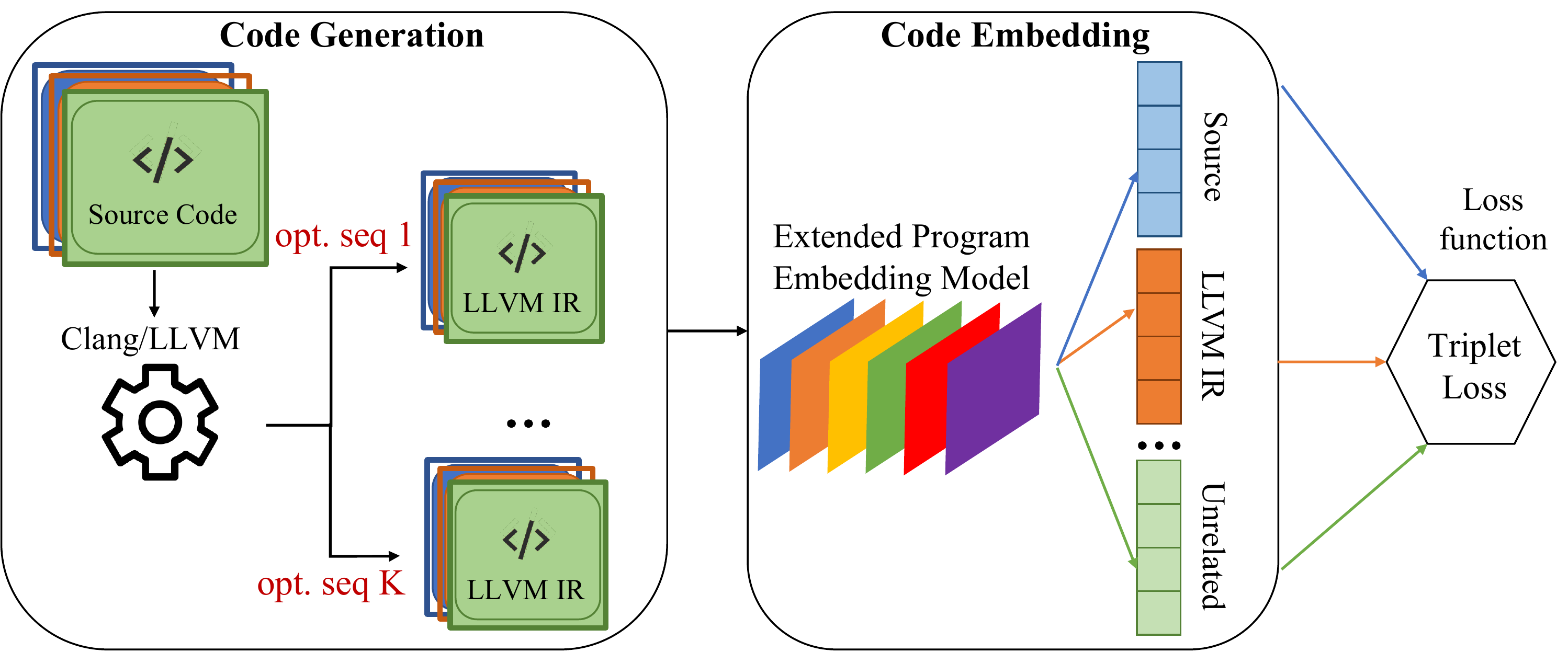}
  \caption{Learning from IR code with Triplet Loss.}
  \label{fig:triplet}
\end{figure}

\F~\ref{fig:triplet} depicts an efficient and general extension over program
embedding models to subsume multiple IR code. As expected, we first extend a
code embedding model $M$ to process LLVM IR. Then, we employ a popular learning
objective, namely triplet loss~\cite{weinberger2009distance}, as the loss
function of $M$.
The triplet, which consists of a positive sample, a negative sample, and an
anchor, is used as the input for triplet loss. An anchor is also a positive
sample, which is initially closer to some negative samples than it is to some
positive samples. The anchor-positive pairs are pulled closer during training,
whereas the anchor-negative pairs are pushed apart. In our setting, a positive
sample represents a program $p$, anchor represents IR code produced from $p$,
and negative samples represent other unrelated source code.

Note that $M$ is not necessarily \codecmr. Other non-trivial source code
embedding models can serve $M$ in this pipeline; see our evaluation on
generalization in \S~\ref{subsec:generalization}. Further, while we adopt
\F~\ref{fig:triplet} to enhance $M$, we clarify that there may exist other
augmentation pipelines. We provide proposals of other pipelines
at~\cite{proposal} for information of interested audiences.

\section{Implementation}
\label{sec:implementation}

\tool\ is written primarily in Python with about 9K lines of code. This
primarily includes our GA pipeline (\S~\ref{sec:design}) and extension of
\codecmr\ (see below). \tool\ is based on LLVM version
11.1.0~\cite{Lattner2004LLVM}. We also tentatively tested LLVM version 7.0,
which works smoothly. \tool\ is built in a fully automated and
``out-of-the-box'' manner. Users only need to configure \tool\ with the path of
their LLVM toolchain. We release \tool\ and data (e.g., augmented models)
at~\cite{website}. Our results can be reproduced using our released artifacts.
We pledge to keep \tool\ up to date to support future study.

\begin{figure}[!ht]
    \centering
    \includegraphics[width=0.65\linewidth]{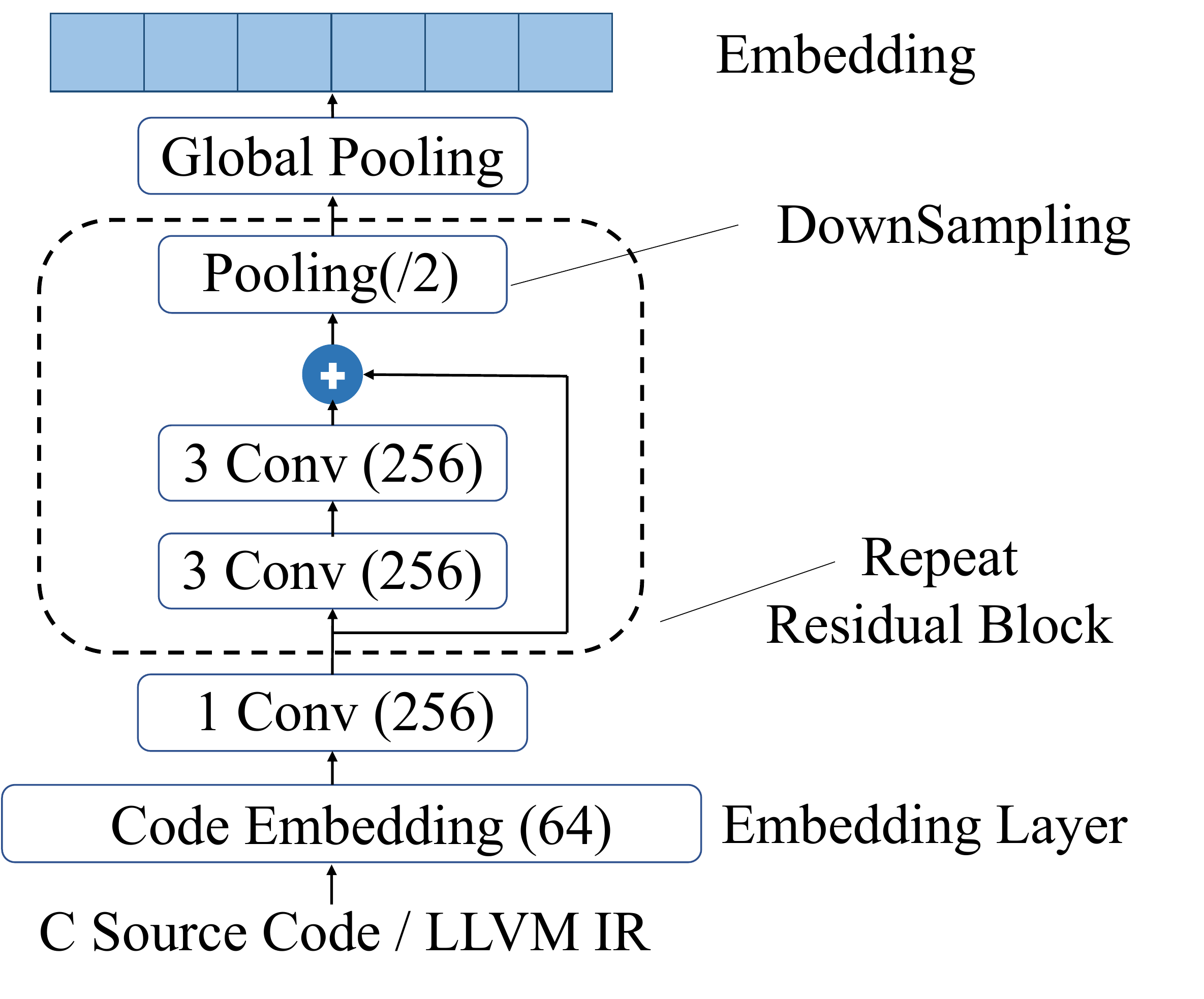}
    \caption{The main structure of CodeCMR.}
    \label{fig:dpcnn}
  \end{figure}

\noindent \textbf{Enhancing \codecmr.}~As stated in \S~\ref{sec:motivation},
\tool is currently implemented with \codecmr~\cite{yu2020codecmr}, which is a
SOTA code embedding model that has been thoroughly tested on real-world C/C++
programs. We find that its code is straightforward to use and of high quality.
We emphasize that \tool\ is orthogonal to the particular code embedding models used.
We assess the generalization of \tool\ using another embedding tool, \ncc, which
directly computes embeddings from LLVM IR; see \S~\ref{subsec:generalization}.
We extended the official version of \codecmr\ to jointly compute embeddings
using C source code and LLVM IR code. We also implement a C/C++ parser based on
treesitter~\cite{treesitter} and a LLVM IR parser (extended from \ncc), as we
need to compare distance of C/C++ and IR code using kernel methods.
\F~\ref{fig:dpcnn} depicts the main network structure of our extended \codecmr.
\codecmr\ is a variant of DPCNN, which has been shown to efficiently represent
\textit{long-range associations} in text. As shown in \F~\ref{fig:dpcnn}, the
key building block, a word-level convolutional neural network (CNN), can be
duplicated until the model is sufficiently deep to capture global input text
representations. Given that (IR) programs are typically lengthy and contain
extensive global information, \codecmr\ exhibits promising accuracy.


%

\begin{table}[!thp]
  \centering
  \scriptsize
  \caption{Augmentation using $K$ collections of optimized IR.}
  \label{tab:merge}
  \resizebox{1.00\linewidth}{!}{
    \begin{tabular}{l|c|c|c|c|c|c|c}
      \hline
      & $k=1$ &$k=2$ &$k=3$ & $k=4$ & $k=5$ &$k=6$ & $k=7$ \\
      \hline
      \textbf{MAP}  & 86.34   & 88.03  & 88.96  &89.71 & 90.16 & \textbf{91.12} & 90.00 \\
      \hline
    \end{tabular}
  }
\end{table}

\noindent \textbf{Tuning $K$.}~Recalling that \tool\ generates $K$ collections
of IR codes by returning the top-$K$ sequences, we now compare the effect of $K$
on learning quality. We ran our experiments using \codecmr\ trained on POJ-104
and measured the MAP accuracy for different $K$ in \T~\ref{tab:merge}. Overall,
although increasing $K$ can continuously extend the training data, the learning
accuracy reached its peak value when $K$=6. We interpret these results to
confirm another important (and intuitive) observation:

\begin{tcolorbox}
  Aligned with data augmentation on natural language or image models, involving
  multiple diverse IR code collections in training datasets augments the learning
  quality of code embedding.
\end{tcolorbox}

We provide samples on~\cite{diverseir} to illustrate how source code can be
compiled into $K$ pieces of diverse IR code. In our evaluation
(\S~\ref{sec:evaluation}), we chose $K$=6 as the default option. However, we
clarify that the best value of $K$, as a hyper-parameter, can be influenced by
both the specific dataset and the neural embedding model. We therefore recommend
users to tune $K$ for their specific learning task.

\section{Evaluation}
\label{sec:evaluation}

Our evaluation aims to answer the following research questions: \textbf{RQ1}:
How does \codecmr, after enhanced by \tool, perform in comparison to other
relevant works on code clone detection?
\textbf{RQ2}: How accurate is the genetic algorithm (GA) adopted by \tool?
\textbf{RQ3}: What are the most important optimization flags and their
characteristics? Does the optimal sequence of flags change on different
datasets?
\textbf{RQ4}: What is the generalization of \tool with respect to other models and different learning algorithms?
\textbf{RQ5}: Can \tool\ still achieve promising augmentation  when only limited source code samples are available?
%
%
Before reporting the evaluation results, we first discuss the evaluation setup
as follows.



\noindent \textbf{Dataset.}~We used the POJ-104~\cite{poj} and GCJ~\cite{gcj}
datasets for our evaluations. \T~\ref{tab:statistics} reports the summary
statistics of these two datasets. As mentioned in \S~\ref{sec:motivation}, the
POJ-104 dataset contains 44,912 C/C++ programs that implement entry-level
programming assignments for 104 different tasks (e.g., merge sort and two sum).
The Google Code Jam (GCJ) is an international programming competition that has
been run by Google since 2008. The GCJ dataset contains the source code from
solutions to GCJ programming challenges. The GCJ dataset is commonly used and
contains 260,901 C/C++ programs.
Compared to POJ, the GCJ files are longer and more numerous. We find that GCJ
files contain complex usage of C macros. As described later in the evaluation,
we found that the more lengthy GCJ code and its relatively complex code
structures had notable impacts on the optimization sequences selected by the GA
procedure. For both datasets, we used the default setting to split them for
training and testing. We did not use their validation splits.
For each dataset, we used \tool\ to select the top-$K$ optimization sequences
retained by the GA process. We then compiled each C source code in the training
datasets into $K$ pieces of LLVM IR code to extend the training datasets.

Our method requires that the training codes be \textit{compilable}. We indeed
explored some other datasets such as Devign~\cite{ZhouLSD019}. However, we found
that many of its cases cannot be compiled. Fixing these issues would have
required considerable manual effort. Another practical concern is \textit{cost};
as soon reported in \textbf{Cost}, training \codecmr\ on GCJ already takes over
90 hours on \textit{16 Tesla V100 GPU cards}. Considering larger datasets is out
of scope for this research project.



\begin{table}[!thp]
  \centering
  \scriptsize
  \caption{Statistics of the dataset used in evaluation.}
  \label{tab:statistics}
\resizebox{0.82\linewidth}{!}{
\begin{tabular}{l|c|c}
\hline
 \textbf{Split} & GCJ &POJ-104 \\
 \hline
  Classes in training data &237  & 64\\
  Programs in training data &238,106  & 28,103\\
  Classes in test data &31  & 24\\
  Programs in test data &22,795  & 10,876\\
  Programs with macro &80,432  & 10\\
  Average lines of C code &71.19  & 35.97\\
  Average lines of LLVM IR code & 1659.50 & 238.51 \\
  
\hline
  \end{tabular}
}
\end{table}

\begin{table*}[!thp] 
  \centering
  \caption{Accuracy of all (augmented) models. For each metrics, we mark
    \colorbox{pptblue}{best models} when training with C source code. We also
    mark the \colorbox{pptgreen3}{best models} when training with C code and
    LLVM IR code optimized following different schemes. \textbf{\tool} denotes
    training \codecmr\ using source code and six collections of LLVM IR
    optimized by sequences formed by \tool.}
  \label{tab:map}
  \resizebox{0.70\linewidth}{!}{
  \begin{tabular}{ccc|cc}
  \hline 
  \multicolumn{1}{c}{\multirow{2}{1.2cm}{Method}} 
  &\multicolumn{2}{c|}{GCJ} & \multicolumn{2}{c}{POJ-104} \\
  \cline{2-5} 
  \multicolumn{1}{c}{} &\multicolumn{1}{c}{MAP@R(\%)} & {AP(\%)} & {MAP@R(\%)} & {AP(\%)}  \\ 
  \hline 
  \ctv\           & 7.76 (-0.79/+0.88) &17.95 (-1.24/+1.76) & 1.90 (-0.43/+0.38) &5.30 (-0.80/+0.60)  \\
  \cts\           & 11.67 (-1.98/+1.73)& 23.09 (-3.24/+2.49)& 3.12 (-0.45/+0.67) &6.43 (-0.37/+0.48)  \\
  \ncc\           & 17.26 (-1.11/+0.57)& 31.56 (-1.11/+1.46)& 39.95 (-2.29/+1.64)& 50.42 (-2.98/+1.61)\\
  \ncci\          & 34.88 (-5.72/+7.63)& 56.12 (-7.63/+9.96)& 54.19 (-3.18/+3.52)& 62.75 (-5.49/+4.42)\\
  \aromad\        & 29.08  & 42.47     & 52.08   &45.99 \\ 
  \aromac\        & 29.67 & 36.21      & 55.12   &55.40 \\ 
  \codecmr\       &64.86(-1.49/+0.72)  & 98.52(-0.16/+0.12) &76.39(-0.55/+1.30) &  77.18(-2.95/+1.92) \\
  \misimg\        &\colorbox{pptblue}{74.90(-1.15/+0.64)} & \colorbox{pptblue}{92.15(-0.97/+0.7)} & 82.45 (-0.61/+0.40) &82.00 (-2.77/+1.65) \\ 
  \codebert\      &68.95(-0.91/+0.37)  & 81.34(-1.29/+0.36)  & \colorbox{pptblue}{82.67(-0.42/+0.33)}   & \colorbox{pptblue}{85.73(-1.14/+2.13)} \\\hline
  \texttt{CodeCMR-O0}      &81.08(-1.03/+0.58)  & 96.31(-0.34/+1.11)  &82.90(-1.24/+0.97)  & 84.95(-2.53/+1.03) \\
  \texttt{CodeCMR-O1}      &83.87(-0.77/+0.24)  & 97.10(-0.27/+0.54)  &83.37(-0.97/+0.31)  & 86.61(-1.35/+0.78) \\
  \texttt{CodeCMR-O2}      &82.60(-0.81/+0.19)  & 96.28(-0.57/+0.27)  &84.29(-1.24/+0.53)  & 85.96(-1.18/+0.91) \\
  \texttt{CodeCMR-O3}      &82.67(-1.13/+0.69)  & 96.77(-0.44/+0.64)  &84.21(-0.43/+0.98)  & 85.06(-0.83/+0.39) \\
  \texttt{CodeCMR-Os}      &85.17(-0.24/+0.38)  & 98.02(-0.31/+0.13)  &83.81(-0.93/+0.24)  & 85.07(-0.72/+1.21) \\
  \texttt{CodeCMR-demix}   &84.93(-1.44/+0.73)  & 98.02(-0.49/+0.31)  &85.14(-0.71/+0.76)  & 88.58(-0.93/+0.44) \\\hline
    \textbf{\tool}  &\colorbox{pptgreen3}{86.48(-1.13/+1.57)} &  \colorbox{pptgreen3}{99.94(-0.07/+0.02)} & \colorbox{pptgreen3}{89.18(-0.33/+0.61)} & \colorbox{pptgreen3}{93.24(-0.21/+0.09)} \\ 
  \hline 
  \end{tabular} 
  }
\end{table*} 

\noindent \textbf{Baseline Models.}~To compare with \codecmr\ augmented by
\tool, we configure seven embedding models, including
\codebert~\cite{feng2020codebert}, \ctv~\cite{alon2019code2vec},
\cts~\cite{alon2018code2seq}, \ncc~\cite{ncc}, \aroma~\cite{luan2019aroma},
\codecmr and \misim~\cite{ye2020misim}. \codecmr\ was introduced in
\S~\ref{sec:motivation}. \codebert, \ncc, and \misim\ were introduced in
\S~\ref{subsec:background-model}.

\ncc\ is a unique and extensible code embedding framework that learns directly
from LLVM IR code. As expected, \ncc\ can be augmented with LLVM IR optimized by
\tool\ (see \S~\ref{subsec:generalization}). When using \ncc, we assessed two
variants, \ncc\ and \texttt{ncc-w/o-inst2vec}. The latter model omits the
standard \texttt{ins2vec} model~\cite{ncc} for IR statement-level embedding and instead
uses a joint learning approach to simultaneously compute and fine-tune the
statement and graph-level embeddings. For \misim, we leveraged its provided
variant, referred to as \misimg, that leverages GNNs in the learning pipeline
and has been shown to outperform other \misim\ variants.

\aroma\ was released by Facebook to facilitate high-speed query matching from a
database of millions of code samples. \aroma\ does \textit{not} perform neural
embedding but instead contains a set of conventional code matching techniques
(pruning, clustering, etc.). We selected \aroma\ for comparison because it is a
SOTA production tool that also features code clustering and similarity analysis.
Hence, \aroma\ and neural embedding tools can be compared on an equivalent
basis, demonstrating the strength of SOTA neural embedding tools, particularly
after augmentation using \tool. The official codebase of \aroma\ provides two
variants, \aromad\ and \aromac. We benchmarked both variants.

\noindent \textbf{Cost.}~Our learning and testing for GA were conducted on a
desktop machine with two Intel Core(TM) i7-8700 CPU and 16GB RAM. The machine
was running Ubuntu 18.04. \tool\ takes averaged 143.52 and 963.41 CPU hours to
finish all 800 iterations of GA procedure for POJ-104 and GCJ, respectively.
Despite the high CPU hours, we clarify that the wall-clock time can be largely
reduced via parallelism. We explored to re-run the GA procedure on a 64-core CPU
server. We report that it takes about 25 wall-clock hours for POJ-104, and about
81 wall-clock hours for GCJ. Setting this parallelism changes about 60 LOC in
\tool; see our codebase at~\cite{website}. When needed, it is also possible to
optimize GA with subsampling for extremely large datasets.

Training embedding models are usually \textit{very costly}. We employ a GPU
server for training for all involved models. The server has two Intel(R) Xeon(R)
Platinum 8255C CPUs operating at 2.50GHz, 384 GB of memory and 16 NVIDIA Tesla
V100 GPU, each with 32GB RAM. The learning rate is 0.001 and the repeat number
of residual blocks is 11; other settings of our extended \codecmr\ are the same
with the standard \codecmr\ setting. In total, over 120 epochs took
approximately 15.9 and 91.3 hours for POJ-104 and GCJ, respectively.

\subsection{Accuracy of \tool}
\label{subsec:eval-overview}

We first answer \textbf{RQ1} using \T~\ref{tab:map}. For neural embedding
models, we launch each experiments for three times and report the average, as
well as the minimum and maximum scores in parentheses. \T~\ref{tab:map} reports
the evaluation results of baseline models in lines 3--11. In accordance with our
research motivation (\S~\ref{sec:motivation}), we also report results using
\codecmr\ augmented with IR code optimized by standard optimization levels (-O0,
-O1, -O2, -O3, -Os). \texttt{CodeCMR-demix} represents training \codecmr\ by
using source code and five sets of IR compiled by all five default optimization
levels.
The last row in \S~\ref{sec:motivation} reports the performance metrics for
\codecmr\ augmented by six collections of LLVM IRs optimized using sequences
generated by \tool. For both the POJ-104 and GCJ datasets, in addition to MAP,
we used AP~\cite{BaezaYatesR99} as the metric. Both metrics are commonly used in
relevant research to assess performance of embedding models. AP stands for
Average Precision, a method combines recall and precision for ranked retrieval
results. 
For both metrics, a higher score indicates better performance.

Our results show that modern embedding models, including \codebert, \codecmr,
and \misimg, can largely outperform conventional code matching tools such as
\aromad\ and \aromac. When learning over C source code, we found that
\codebert\ was the best performing model for the POJ-104 dataset, whereas
\misimg\ delivered the best performance for the GCJ dataset. In contrast,
\codecmr\ performed less well than either of the SOTA models across all of the
evaluated metrics. \ctv\ and \cts\ shows relatively lower accuracy compared with
others. Since we run each evaluation three times, we find that their accuracy
scores are unstable. Such observation is also consistently reported in previous
works~\cite{ye2020misim}. Nevertheless, even the ``peak'' accuracy scores of
them are still much lower than that of the SOTA models.

When learning from IR optimized using standard optimization levels,
\codecmr\ outperformed SOTA model MAP scores by more than 10\% on the GCJ
dataset. Evaluation of this form of \codecmr\ training on the POJ-104 dataset
showed consistently promising enhancement relative to the SOTA models in most
cases. Also, comparing with augmenting \codecmr\ with one collection of
optimized IR code, the \texttt{CodeCMR-demix} setting shows (slightly) better
performance, particularly for the POJ-104 setting. This also reveals the
strength of training with multiple diverse sets of IR code.

We found that \codecmr, when augmented by findings of \tool\ (the last row of
\T~\ref{tab:map}), constantly and notably outperformed all the other settings.
We interpret the evaluation results as highly encouraging, showing that \tool\
can generate high-quality LLVM IR code that enhances \codecmr\  to significantly
outperform the SOTA models (\codebert\ and \misimg) on all metrics. Again, we
note that \tool\ is not limited to enhance \codecmr: we present evaluation of
enhancing \ncc\ in \S~\ref{subsec:generalization}.

\begin{figure}[!ht]
  \centering \includegraphics[width=1.00\linewidth]{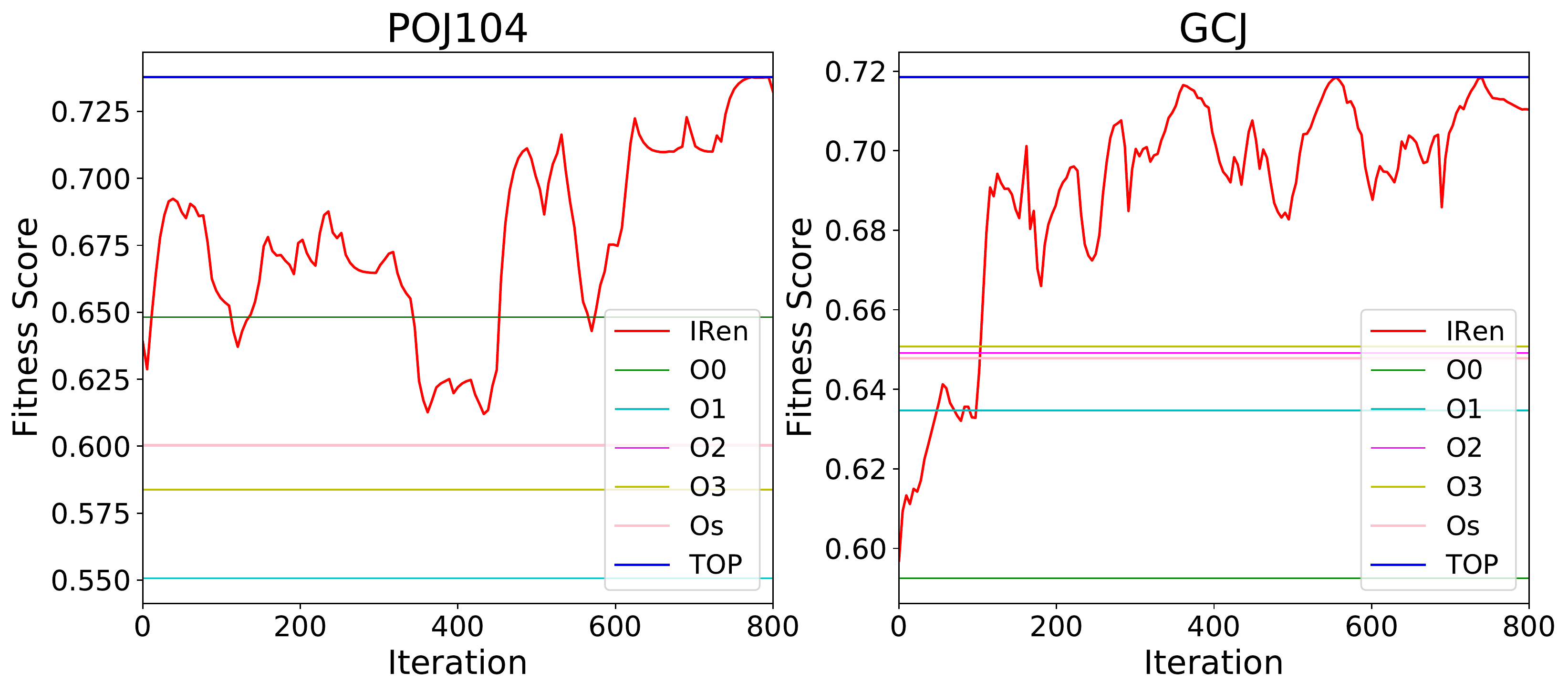}
  \caption{Smoothed fitness score increases over 800 iterations.}
  \label{fig:reward}
\end{figure}

\subsection{Fitness Function}
\label{subsec:rq2}

\textbf{RQ2} assesses the efficiency of our fitness function.
\F~\ref{fig:reward} reports the fitness score increases from all 800
\tool\ iterations across each GA campaign. The test cases, despite their diverse
functionality, manifested encouraging and consistent trends during optimization
searching. The fitness scores kept increasing and were seen to reach saturation
performance after around 410 to 600 iterations. We interpret that under the
guidance of our fitness function, \tool\ can find well-performing sequences for
both datasets.

\begin{figure}[!ht]
  \centering \includegraphics[width=1.00\linewidth]{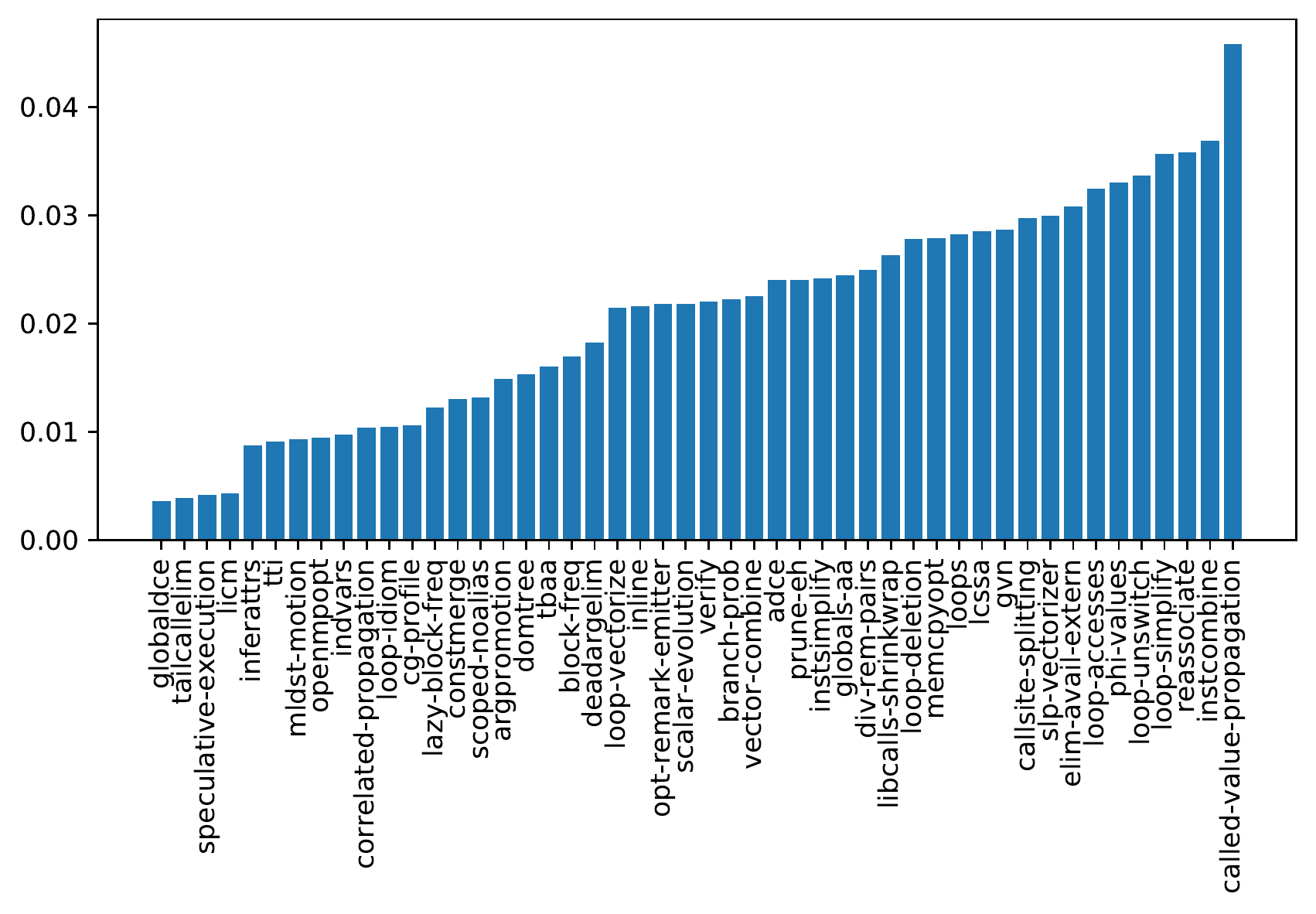}
  \caption{Ordered contributions of each optimization flag.}
  \label{fig:FlagsAna}
\end{figure}

\subsection{Potency of Optimization Flags}
\label{subsec:rq3}

This section answers \textbf{RQ3} by measuring the potency of selected
optimization flags (selected flags are fully listed at~\cite{website}). We
report that for the POJ-104 dataset, the top-$1$ sequence $S$ having the highest
fitness score contains 49 flags. To measure their contribution, we first train
\codecmr\ using C source code and LLVM IR optimized using sequence $S$ and
record the baseline accuracy as $acc$. Then, we iteratively discard one
optimization flag $f$ from $S$ and measure the augmentation effectiveness of
using the remaining sequence with 48 flags. The accuracy drop reveals the
contribution of flag $f$.

\F~\ref{fig:FlagsAna} orders the contribution of each flag in $S$. Overall, we
interpret that no ``dominating'' optimization flags are found in this
evaluation. In other words, we interpret that all these 49 flags manifest
reasonable contribution to the model augmentation, and the top-10 flags
contributes in total 34.38\%. We thus make our first important observation
w.r.t.~\textbf{RQ3}:

\begin{tcolorbox}
  Instead of identifying one or few \textit{dominating flags} that
significantly contribute to enhancing code embedding, it is rather the formed
sequence of optimization flags that is important.
\end{tcolorbox}

This evaluation shows that a sequence of flags works together to produce
high-quality IR, instead of one or a few ``franchise players'' that can largely
outperform other flags. In other words, the GA process conducted by \tool\ is
\textit{critical} to this research, because it offers a general way to construct
such a sequence with modest cost.

%
%

We now consider the characteristics of the ten highest potency flags. We put
these flags into three categories as follows:

\noindent \textbf{Simplify an IR Statement.}~Optimization flags, including
\texttt{-dce}, \texttt{-early-cse}, \texttt{-reassociate}, \texttt{-bdce} and
\texttt{-loop-deletion}, simplify IR statements via various data flow or control
flow analysis methods. For instance, \texttt{-early-cse} regulates IR statements
by eliminating common subexpression eliminations, and \texttt{-reassociate}
reassociates commutative expressions to support better constant propagation. In
all, these optimization can make two syntactically distinct pieces of source
code more similar in IR.

\noindent \textbf{Make IR Statement Sequences Closer to Source Code.}~Flags,
including \texttt{-mem2reg}, \texttt{-instcombine}, and \texttt{-dse}, can
simplify the compiled IR code, making it \textit{more similar to the source
  code}. For instance, \texttt{-mem2reg} promotes memory references to be
register references. This prunes verbose memory usage in IR code and generates
IR code similar with source code in terms of memory reference. Other flags, such
as \texttt{-instcombine} and \texttt{-dse}, combine and simplify instructions to
form fewer and more succinct instruction sequences that are generally closer to
source code.

\noindent \textbf{Simplify CFG.}~Optimization flags, including
\texttt{-break-crit-edges}, \texttt{-simplifycfg}, and \texttt{-loop-rotate},
perform more holistic transformations to simplify the IR code CFGs. For
instance, \texttt{-simplifycfg} performs dead code elimination and basic block
merging by eliminating useless basic blocks and their associated LLVM PHI nodes.
By regulating CFGs, these optimizations deliver similar IRs from two
semantically similar but syntactically different source codes.

Our analysis identified numerous optimization flags that significantly improved
the training IRs for embedding learning. This is intuitive, given that they
launch transformation from different granularities. More importantly, we make
the following observation to characterize important optimization flags:

\begin{tcolorbox}
Optimization passes that \textit{simplify and regulate IR code}, either at the
IR statement level or the CFG level, are generally desirable.
\end{tcolorbox}

Many of these flags are often employed as cleanup passes to run after compiler
inter- or intra-procedural optimizations.

\subsection*{Optimization Passes on GCJ}

We further analyze the top-$1$ optimization sequences found by \tool\ for the
GCJ dataset. This top-1 sequence contains 50 flags. Given that the top-1
sequence found over POJ-104 contains 49 flags, we further measure the agreement
of these two sequences by counting the number of flags appeared in both
sequences. These two sequences agree on 28 flags. The top-3 flags in the POJ-104
sequence all exist in the intersection set, and five of the top-10 flags in the
POJ-104 sequence exist in the intersection set. With respect to the
(dis)agreement, we conduct a manual investigation and summarize our findings as
follows:

\noindent \textbf{Agreement.}~We found that these 29 overlapping flags
primarily serve the purpose of simplifying and regulating IR code in different
ways. 
For instance, \texttt{-reassociate} and
\texttt{-deadargelim} simplify IR statements and delete dead arguments. 
The rest overlapping flags are used as utilities for other passes (e.g., 
\texttt{-lcssa} serves loop-oriented optimizations) or  
for analysis purposes (e.g., \texttt{-block-freq}).
Overall,
we interpret that code cleanup and regulation are generally applicable to
enhancement of learning quality.

\noindent \textbf{Disagreement.}~Given that POJ-104 test cases are relatively
succinct, we find that \tool\ tended not to select flags that focus on shrinking
the code size. In contrast, GCJ contains much lengthy C/C++ code, whose derived
LLVM IR code is even more lengthy. Hence, \tool\ adaptively prioritizes more
flags to reduce the size. Overall, we find that whether \tool\ inclines to
``shrink'' code size is dataset-dependent. IR compiled from POJ is generally
shorter than that of GCJ; therefore, it has fewer OOV issues and the need for
shrinking is less frequent. GCJ has lengthy IR and more OOV IR statements; it is
demanding to shrink IR to avoid OOV. 
In addition, GCJ features more floating number related programming tasks, and
accordingly, floating number related flags, such as \texttt{-float2int}, are
involved to turn floating numbers into integers and effectively reduce the \#OOV
cases. In contrast, POJ-104 dataset does not primarily involve floating
number-related computations.

Overall, we interpret these results as promising: we found that over half of the
optimization flags selected by \tool (29; $\sfrac{29}{49}=59.2\%$ of all flags
selected over POJ-104) were selected across two datasets of different complexity
without using any pre-knowledge or in-depth program analysis. These overlapping
flags further highlight the importance of cleaning up and regulating IR code to
make neural embedding models more robust. Moreover, the 42.9\% disagreement, to
some extent, shows that \tool\ enables the selection of more diverse flags
adaptive to different datasets.

\begin{table}[!thp]
  \centering
  \scriptsize
  \caption{Augment \ncc\ over the POJ-104 dataset.}
  \label{tab:nccaug}
\resizebox{0.65\linewidth}{!}{
\begin{tabular}{l|c}
\hline
      Model & MAP@R(\%)  \\
 \hline
      \ncc\                 &   39.95 \\
      \ncc-random           &   40.34 \\
      \ncc-\textsc{IRGen}   &   56.07 \\
      \ncci\                &   54.19 \\
      \ncci-random          &   55.10 \\
      \ncci-\textsc{IRGen}  &   60.46 \\
      
      \hline
\end{tabular}
}
\end{table}

\subsection{Generalization}
\label{subsec:generalization}

As aforementioned, augmentation (including the fitness function; see
\S~\ref{subsec:fitness}) delivered by \tool\ is \textit{independent} from
particular embedding model design. To answer \textbf{RQ4}, we demonstrate the
generalization of \tool\ by augmenting another popular neural embedding model,
\ncc. As previously described, \ncc performs embedding over LLVM IR code.
Therefore, we did not need to change the implementation of \ncc.
The \ncc\ augmentation evaluation results are reported in \T~\ref{tab:nccaug}.
To compare with optimization sequences formed by \tool, we also prepare a
``baseline'', denoting a sequence containing randomly selected 49 optimization
flags. These two baseline results are reported in third and sixth rows.

As expected, augmentation notably improved the quality of \ncc\ and \ncci.
Particularly, the MAP score of the latter model is improved to 60.46\%, which is
even higher than the scores achieved by two variants of Aroma. In contrast, we
find that two random schemes show negligible enhancement. Overall, this
evaluation indicates that \tool delivers general augmentation for neural code
embedding models, without consideration of the specific model designs.
%

\begin{figure}[!ht]
  \centering \includegraphics[width=1.00\linewidth]{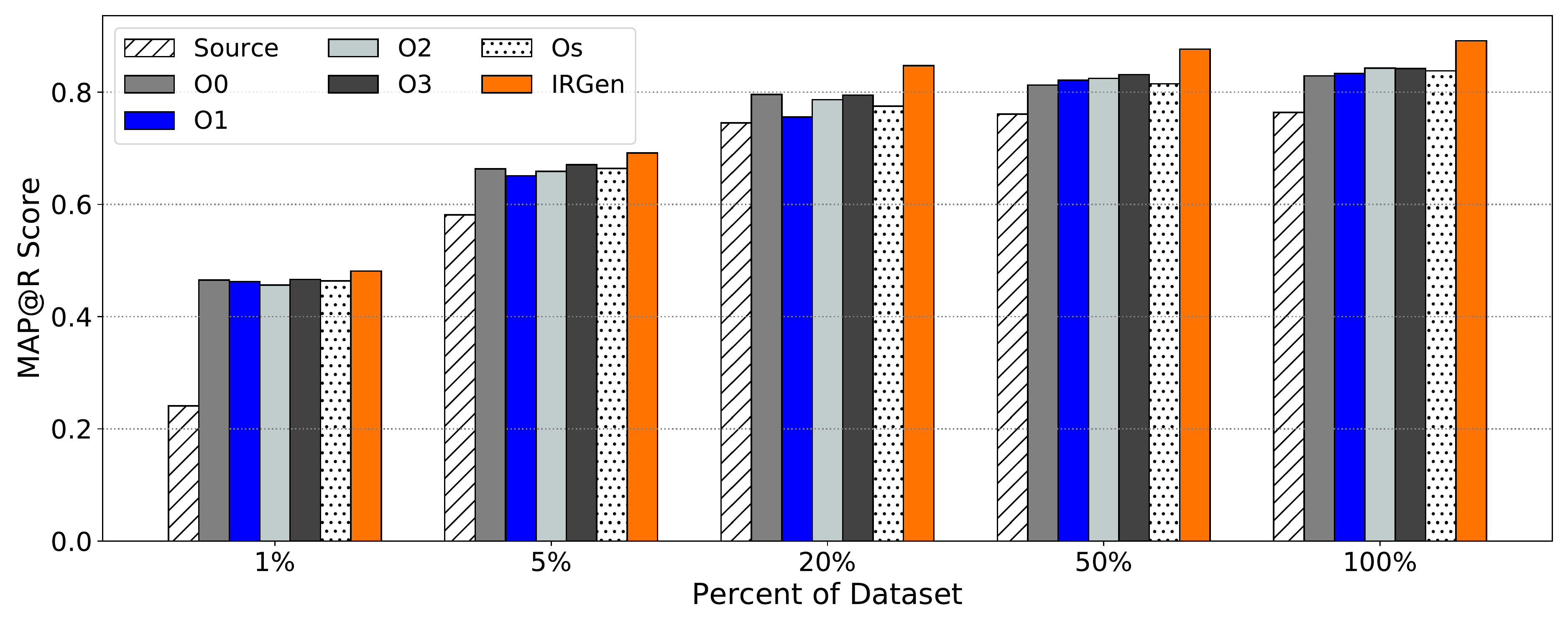}
  \caption{Performance with different size of training data.}
  \label{fig:small}
\end{figure}

\subsection{Augmentation with Small Training Data}
\label{subsec:SmallDataset}

\textbf{RQ5} considers a scenario where program embedding is inhibited by a
limited amount of training data. We argue that this is a common situation, such
as in vulnerability analysis where only limited vulnerable code samples are
available. In these situations, we anticipate the feasibility and usefulness of
extending training dataset and augmenting embedding models using optimized IR
code.

\F~\ref{fig:small} presents the evaluation results of augmenting a small
training dataset. Specifically, we randomly selected 1\% of the C source codes
from the POJ-104 training data to train \codecmr\ and measured the resulting
embedding accuracy, which was quite low (the first bar in \F~\ref{fig:small};
MAP = 24.08\%). However, after using different standard optimization levels and
optimization sequences selected by \tool, the MAP accuracy increased to over
40\%, almost \textit{doubling} the original MAP score. In addition to an extreme
1\% sample test, we also randomly selected 5\%, 20\%, and 50\% of the POJ-104
training data and re-trained \codecmr. As shown in \F~\ref{fig:small}, we
further augmented the model for each subset by using the same standard
optimization flags and flags selected by \tool. We consistently achieved
promising augmentation results. In particular, optimization flags found by
\tool\ outperformed all other augmentation schemes in all of these small
training data settings. These intuitive and highly promising evaluation results
indicate that IR code can be leveraged to significantly and reliably augment
code embedding, particularly when only very limited data are available.
Comparing with standard optimization levels, optimization sequences selected by
\tool\ can result in even higher enhancement.

  

\section{Discussion}
\label{sec:discussion}

\noindent \textbf{Generalizability of Downstream Applications.}~This work
focuses on a representative downstream task of code embedding --- code clone
detection. To clarify the generalizability behind ``code clone'': program
embeddings can be used as the basis of a variety of downstream tasks like
malware clustering, vulnerability detection, and code plagiarism detection.
Holistically, many of these applications are based on deciding two software's
\textit{similarity}. Therefore, we view ``code clone'' detection as a core basis
to assess those applications. Nevertheless, the augmentation
pipeline of \tool\ is generally \textit{orthogonal} to a particular downstream task. We
leave it as one future work to benchmark the augmentation capability offered by
\tool\ toward other important downstream applications, such as 
vulnerability detection and program repair. We envision to have consistently
promising observations.

\noindent \textbf{Conflicts Between Optimization Flags.}~\gcc\ documents a set
of constraints between optimization passes~\cite{gccopt} wherein using two
conflicting passes can result in compilation errors.
However, to our knowledge, the LLVM compiler framework does not explicitly
document any ``conflicting'' flags.
We are also not aware of any compilation errors caused by using two conflicting
LLVM passes. In cases where one flag has conflicts with other flags, such
information can be explicitly encoded as constraints to validate generated
optimization sequences.

\noindent \textbf{Other Learning Methodologies.}~Production compiler frameworks
like \gcc\ and LLVM provide considerable optimization flags, forming a large
search space in our research scenario. \tool\ constitutes a GA-based approach to
search for (near-)optimal optimization sequences. Our empirical evaluation shows
that the proposed learning process is \textit{sufficient} to identify
high-performing optimization sequences. We note that there are more advanced
(evolutionary) optimization algorithms available. In particular, the two fitness
objectives could have been optimized separately (i.e., with multi-objective
optimization).
Also, from a holistic view, searching for optimization sequences is a Markov
Decision Process (MDP). Complex MDPs (e.g., auto-driving) can be likely
addressed with reinforcement learning (RL) techniques. Future work may explore
using advanced deep RL models, which have achieved prominent success in solving
real-world challenges in autonomous
driving~\cite{dosovitskiy2017carla,toromanoff2020end} and video
games~\cite{mnih2013playing}.

\section{Related Work}
\label{sec:related}

We reviewed program embedding from various perspective in
\S~\ref{sec:preliminary}. The development of \tool\ was inspired by existing works in
search-based software engineering~\cite{harman2012search,harman2001search} and
search-based iterative compilation techniques~\cite{knijnenburg2001iterative,
cooper2002adaptive, kulkarni2004fast, chen2010evaluating, ansel2014opentuner}.
In general, many tasks in software engineering require exploration of (optimal)
solutions under a range of constraints and trade-offs between resources and requirements. To
this end, metaheuristic algorithms, such as local search, simulated annealing
(SA)~\cite{kirkpatrick1983optimization}, genetic algorithms
(GAs)~\cite{whitley1994genetic}, and hill climbing (HC)~\cite{selman2006hill}
are frequently used to address these challenges. Typical applications include testing and
debugging~\cite{mcminn2011search,harman2015achievements,mcminn2004search},
verification~\cite{alba2007acohg,alba2007ant,alba2007finding,chicano2008finding,godefroid1997model},
maintenance~\cite{o2008search,simons2015search}, and software
hardening~\cite{ghaith2012improving,ghaith2012improving,thome2014search}.

A line of relevant and actively-developed research augments software obfuscation
by combining obfuscation passes. Liu et al.~\cite{liu2017stochastic} search for
a sequence of obfuscation passes to maximize obfuscation effectiveness (and thus
make software more secure). Amoeba~\cite{wang2017composite} empirically
demonstrated that combining obfuscation passes, though enhancing obfuscation
potency, often carries high costs. Wang et al.~\cite{wang2020generating} trained
a reinforcement learning model to explore optimal obfuscation combinations by
taking both cost and effectiveness into account.
BinTuner~\cite{ren2021unleashing} uses a guided stochastic algorithm to explore
how combinations of compiler optimization passes can obfuscate software.

\section{Conclusion}
\label{sec:conclusion}

Existing neural program embedding methods are generally limited to processing of
program source code. We present simple, yet effective, strategies to improve
embedding quality by augmenting training datasets with compiler-generated IR
code. In addition to use of default compiler optimization levels, we present
\tool, a search-based framework to find customized optimization sequences that
produce IRs that can substantially improve the quality of learned embeddings. In
evaluation, these models outperformed others trained using only source code or
IR generated with default optimization combinations. Our study provides insights
and guidance for users aiming to generate higher quality code embeddings.

\section*{Acknowledgements}

This work was supported in part by CCF-Tencent Open Research Fund.
We are grateful to the anonymous reviewers for their valuable comments. 

\bibliographystyle{ACM-Reference-Format}
\bibliography{bib/machine-learning,bib/timing,bib/sidechannel,bib/analysis,bib/ref,bib/bsgx,bib/testing-cv,bib/cv,bib/similarity}

\end{document}